\documentclass[prb,aps,amsmath,twocolumn,amsfonts,superscriptaddress,floatfix,showpacs]{revtex4}

\usepackage{graphicx}

\def\be{\begin{equation}}
\def\ee{\end{equation}}

\newcommand{\corr}[1]{\langle #1\rangle}
\def\str{\mathop{\rm str}}
\renewcommand{\Re}{\mathop{\rm Re}}
\renewcommand{\Im}{\mathop{\rm Im}}
\newcommand{\Reg}{\mathop{\rm Reg}}
\def\erf{\mathop{\rm erf}}
\def\br{\mathbf{r}}
\def\diag{\mathop{\rm diag}}
\def\ZO{\mathop{\hat{\rm Z}}\nolimits_{\lambda^*+1}}

\begin{document}

\title{Correlations of the local density of states
in quasi-one-dimensional wires}

\author{D. A. Ivanov}
\affiliation{Institute of Theoretical Physics,
Ecole Polytechnique F\'ed\'erale de Lausanne (EPFL),
CH-1015 Lausanne, Switzerland}

\author{P. M. Ostrovsky}
\affiliation{Institut f\"ur Nanotechnologie, Forschungszentrum Karlsruhe,
76021 Karlsruhe, Germany}
\affiliation{Landau Institute for Theoretical Physics, 142432 Chernogolovka,
Moscow region, Russia}

\author{M. A. Skvortsov}
\affiliation{Landau Institute for Theoretical Physics, 142432 Chernogolovka,
Moscow region, Russia}
\affiliation{Moscow Institute of Physics and Technology, 141700 Moscow, Russia}

\date{February 5, 2009}

\begin{abstract}
We report a calculation of the correlation function of the local density of states
in a disordered quasi-one-dimensional wire
in the unitary symmetry class
at a small energy difference.
Using an expression from the supersymmetric sigma-model,
we obtain the full dependence of the two-point correlation function
on the distance between the points.
In the limit of zero energy difference,
our calculation reproduces the statistics
of a single localized wave function.
At logarithmically large distances of the order of the Mott scale,
we obtain a reentrant behavior similar to that in strictly one-dimensional chains.
\end{abstract}

\pacs{
73.20.Fz, 
73.21.Hb, 
73.22.Dj  
}

\maketitle

\section{Introduction}

Anderson localization in quasi-one-dimensional (Q1D) disordered systems
is known as a broad universality class for various problems
of contemporary condensed matter physics and of quantum chaos.
\cite{Efetov-book,IWZ,FM91,Fishman,AZ96,SBK04} Quasi-one-dimensional wires are
characterized by a large number $N\gg1$ of transverse channels. Starting from
the mean free path scale $l$, electrons propagate diffusively and eventually
get localized\cite{Thouless77} with the localization length
\cite{Dorokhov} $\xi\sim Nl \gg l$. Quantum diffusion in the broad range of
length scales is the key feature of Q1D systems distinguishing
them from strictly one-dimensional (1D) disordered chains, where localization
occurs immediately at the scale of the mean free path.
\cite{GertsenshteinVasilev1959,MottTwose61}
It is this intermediate diffusive
regime that makes studying localization in Q1D systems a very challenging
problem.

In order to describe the nonperturbative regime of strong
localization, a rather sophisticated mathematical technique
is required, both in the 1D and Q1D cases.
For the strictly 1D geometry, an appropriate technique
has been elaborated by Berezinsky,
\cite{Berezinsky73,BerezinskyGorkov79}
and recently it was translated into a field-theoretic language
in Ref.~\onlinecite{KravtsovOssipov}.
In a Q1D geometry, where localization takes place in the process
of quantum diffusion, the diffusive supersymmetric $\sigma$-model
introduced by Efetov\cite{Efetov83,Efetov-book} is a powerful
tool to describe the system.

Despite the physical difference between the two models,
localization in the 1D and Q1D geometries often look similar.
This analogy is most pronounced in the behavior of
a \emph{single}\/ localized wave function: The statistics of the smooth
wave function envelopes are precisely the same
in the 1D \cite{Gogolin,LGP82,Kolokolov1995}
and Q1D \cite{Mirlin-JMP-1997} geometries,
while the local statistics of the rapidly oscillating components
involves the plane-wave (1D)
or random-matrix (Q1D) moments.\cite{Mirlin-review}

On the other hand, comparison between the quantities involving
\emph{different}\/ wave functions in the two models is problematic
since almost nothing is known about the eigenfunction correlations
in the Q1D geometry in the most physically interesting limit
of sufficiently small energy difference $\omega$.

At the same time, quite a lot is known about eigenfunction
correlations in the strictly 1D geometry under similar conditions.
One of the most important results is the expression
for the dissipative low-frequency conductivity which
follows the Mott-Berezinsky law:\cite{Mott68, Berezinsky73}
$\Re\sigma(\omega) \propto \omega^2\ln^2(1/\omega\tau)$, where $\tau$ is the
elastic time.
Another result concerns the behavior of the correlation function
\be
  R(\omega; \br_1,\br_2)
  = \nu^{-2} \langle
      \rho_{\varepsilon}(\mathbf{r}_1)
      \rho_{\varepsilon+\omega}(\mathbf{r}_2)
    \rangle
\label{R-def}
\ee
of the local density of states (LDOS)
$\rho_\varepsilon(\mathbf{r}) = \sum_n |\psi_n(\mathbf{r})|^2
\delta(\varepsilon-\varepsilon_n)$, where $\nu$ is the average density
of states (for spinless particles). The dependence of $R(\omega; \br_1,\br_2)$
on $|\br_1-\br_2|$ has been addressed for 1D geometry
by Gor'kov, Dorokhov and Prigara \cite{GDP83}.
They have identified that in the limit of small energy separation,
$\omega\tau\ll1$, the spatial behavior of the LDOS correlator
is determined by two scales: the localization length,
$\xi_{\text{1D}} \sim l$, and the Mott scale,
$L_M^{\text{1D}} \sim \xi_{\text{1D}}\ln(1/\omega\tau) \gg \xi_{\text{1D}}$.
The eigenstates
are uncorrelated at $r\gg L_M^{\text{1D}}$ and exhibit nearly
perfect level repulsion at $\xi_{\text{1D}} \ll r \ll L_M^{\text{1D}}$,
which gets reduced at $r \ll \xi_{\text{1D}}$.
Such a behavior can be qualitatively understood \cite{SI87}
using the Mott's picture \cite{Mott1970} of two nearly degenerate
localized states separated by the distance $L_M^{\text{1D}}$
which hybridize to form a pair of eigenstates effectively contributing
to $R(\omega;\br_1,\br_2)$.

In the Q1D geometry, neither the derivation of the low-frequency
conductivity nor the full dependence of the LDOS correlation function
have been reported. Though the one-dimensional diffusive supersymmetric
$\sigma$-model has been reformulated by Efetov and Larkin \cite{EL83}
as a quantum-mechanical problem, the resulting set of differential equations
is still too complicated and resisting a naive perturbative expansion
in the limit $\omega\to0$.

Recently, a significant progress in solving this effective quantum
mechanics has been achieved by two of the authors, \cite{SO07}
who found the exact zero mode of the transfer-matrix Hamiltonian
for the unitary symmetry class. As a result, a non-perturbative
expression for the short-scale (at distances $|\br_1-\br_2|\ll\xi$)
behavior of the LDOS correlator (\ref{R-def}) has been derived
for arbitrary frequencies $\omega$. For the first time,
the difference in the correlations of the localized states in the 1D and Q1D
geometries has been demonstrated.

The purpose of the present work is to develop a regular approach
to the small-$\omega$ expansion of the one-dimensional diffusive
supersymmetric $\sigma$-model. We will assume the unitary symmetry
class and employ the exact knowledge of the zero mode obtained
in Ref.~\onlinecite{SO07}.
Developing an advanced perturbation theory which treats
both powers and logarithms of $\omega$,
we will derive the
expression [Eq.~(\ref{A-M})] for the LDOS correlation function
at arbitrary distances $|\br_1-\br_2|$ between the observation points
and analyze its behavior in various asymptotic regions.

The LDOS correlation function (\ref{R-def}) can generally
be written as \cite{SO07}
\be
  R(\omega; \br_1, \br_2)
  =
  1
  + A(\omega,t)
  + k(\br_1,\br_2) B(\omega).
\label{R-AB}
\ee
Here the factor $k(\mathbf{r}_1, \mathbf{r}_2) = \langle \Im
G^R(\mathbf{r}_1,\mathbf{r}_2)\rangle^2/(\pi \nu)^2$ accounts for
short-scale Friedel oscillations.
\cite{BlanterMirlin97, Mirlin-review}
It is equal to 1 at coincident points and exponentially decays
as $|\br_1-\br_2|$ exceeds the mean free path.
The function $A(\omega,t)$ describes long-range correlations,
with
\be
  t = x/\xi
\label{t-def}
\ee
measuring the distance $x=x_1-x_2$ between the observation points
in units of the localization length,
\be
  \xi = 2\pi\nu_1D ,
\label{xi}
\ee
where $D$ is the diffusion coefficient
and $\nu_1=\nu S$ is the one-dimensional
density of states ($S$ is the wire cross section).

The functions $A(\omega,0)$ and $B(\omega)$ which determine
local correlations at $x\ll\xi$ have been
calculated in Ref.~\onlinecite{SO07}:
\begin{align}
  A(\omega,0)
   &= \frac{4}{3} \Re \big[ \kappa^2 \big( I_1^2 - I_0 I_2 \big)
       \big( K_1^2 - K_0 K_2 \big) - I_1^2 K_1^2
     \big],
\label{Aw0}
\\
  B(\omega)
   &= \frac{4}{3} \Re \big(
       I_1 K_1 + 2 I_2 K_0
     \big),
\label{B}
\end{align}
where $\kappa = \sqrt{-4i\omega/\Delta_\xi}$
and the argument $\kappa$ of the modified
Bessel functions is suppressed.

\begin{figure}
\includegraphics{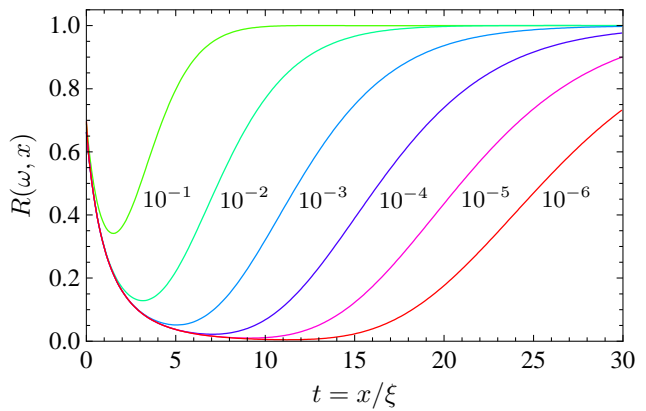}
\caption{(color online). Dependence of the LDOS correlation function $R(\omega,x)$
on $t=x/\xi$ for various values of $\omega/\Delta_\xi$ shown on the graph.}
\label{F:Mott}
\end{figure}

At large separations exceeding the mean free path,
$x\gg l$, the function $k(\br_1,\br_2)$ exponentially
decays and Eq.~(\ref{R-AB}) simplifies to
\be
  R(\omega, x)
  =
  1
  + A(\omega,t)
  .
\label{R-A}
\ee
The behavior of $R(\omega, x)$ as a function of $t=x/\xi$
calculated with the help of the general expression (\ref{A-M})
is shown in Fig.~\ref{F:Mott}. Different curves on the
graph correspond to different values of the ratio
$\omega/\Delta_\xi$, where the level spacing
within the localization length,
\be
  \Delta_\xi = D/\xi^2,
\ee
is the natural frequency scale in the localization problem.

One can clearly see that in the deeply localized regime,
$\omega\ll\Delta_\xi$, the behavior of $R(\omega,x)$ is similar
to the behavior in the strictly 1D geometry: \cite{GDP83,SI87}
On increasing $x$, the function $R(\omega,x)$ first decays
at the localization length and then
reaches the uncorrelated value $R=1$ at the Mott scale
\be
  L_M = 2\xi\ln(\Delta_\xi/\omega) .
\label{L-Mott}
\ee
The function $R(\omega,x)$ has a minimum of the order
of $(\omega/\Delta_\xi)^{1/4}$ at $x\sim L_M/2$.

The paper is organized as follows.
In Sec.~\ref{S:SUSY},
starting with the supersymmetric $\sigma$-model formalism,
we give the expression of the LDOS correlation function
in terms of a matrix element of the evolution operator with the
effective Hamiltonian. The bosonic and fermionic eigensystems
of this Hamiltonian are perturbatively constructed in
Secs.~\ref{S:Bosonic} and \ref{S:Fermionic}, respectively.
In Sec.~\ref{S:result} we give the general expression for the
LDOS correlator and evaluate it for arbitrary $x$ and small $\omega$.
The opposite limit of small $x$ and arbitrary $\omega$ is considered
in Sec.~\ref{S:small-omega}. Section \ref{S:Discussion} is devoted
to the discussion of results and comparison between the 1D and Q1D
localization.
Finally, a large portion of technical details is relegated to several
Appendices.

\section{Supersymmetric sigma-model and general
expression for the LDOS correlation function}
\label{S:SUSY}

An efficient analytical tool for studying disordered metallic systems ---
nonlinear supersymmetric $\sigma$-model --- was proposed in the works of Efetov
\cite{Efetov83,Efetov-book}.
This approach is based on the effective action
\be
\label{S[Q]}
  S[Q]
  =
  -
  \frac{\pi\nu_1}{4}
  \str
  \int dx
  \biggl[
    D \left( \frac{dQ}{dx} \right)^2
  + 2i\omega \Lambda Q
  \biggr]
\ee
for slow diffusion modes
described by the $4 \times 4$ supermatrix $Q$ acting in the direct
product of the Fermi--Bose (FB) and Retarded--Advanced (RA) spaces
(we assume the unitary symmetry class). The $Q$ matrix is subject
to the nonlinear constraint $Q^2=1$. In Eq.~(\ref{S[Q]}),
$\Lambda=\diag(1,-1)$ the matrix in the RA space.

The one-dimensional $\sigma$-model (\ref{S[Q]}) can be mapped \cite{EL83}
onto a quantum mechanics of the $Q$ matrix, with the coordinate $x$
along the wire playing the role of time. In that language, the functional
integral with the action (\ref{S[Q]}) describes evolution
of the wave function $\Psi(Q)$ under the action of an appropriate
Hamiltonian ${\cal H}$:
\be
  \frac{\partial}{\partial t} \Psi(Q)
  =
  -
  2 \, {\cal H} \, \Psi(Q) ,
\ee
where $t$ is the dimensionless ``time'' (\ref{t-def})
measured in units of the localization length.
In the unitary symmetry class, the $Q$ matrix is generally
parameterized by four commuting and four Grassmann variables.
However in many cases it suffices to consider only the
``radially-symmetric'' (``singlet'') wave
functions $\Psi(Q)=\Psi(\lambda_B,\lambda_F)$
which depend only on Efetov's coordinates
$\lambda_B \in [1,\infty)$ and $\lambda_F \in [-1,1]$,
parameterizing the non-compact BB
and compact FF sectors.\cite{Efetov83,Efetov-book}
On the space of such functions, the Hamiltonian
${\cal H}$ acquires the form\cite{EL83,Efetov-book}
\begin{multline}
  H = -\frac{(\lambda_B-\lambda_F)^2}2
  \biggl[
    \frac{\partial}{\partial\lambda_F}
    \frac{1-\lambda_F^2}{(\lambda_B-\lambda_F)^2}
    \frac{\partial}{\partial\lambda_F}
\\ {}
  +
    \frac{\partial}{\partial\lambda_B}
    \frac{\lambda_B^2-1}{(\lambda_B-\lambda_F)^2}
    \frac{\partial}{\partial\lambda_B}
  \biggr]
  + \Omega (\lambda_B-\lambda_F) ,
\label{H}
\end{multline}
where we have introduced the dimensionless frequency
\be
  \Omega
  =
  - \frac{i\omega}{4\Delta_\xi} .
\label{Omega-def}
\ee

At $\omega=0$, the Hamiltonian (\ref{H}) reduces
to the Laplace operator on the space of $Q$-matrices.
This operator is known \cite{Zirnbauer92,MMZ94} to have a discrete spectrum
in the FF sector and a continuous spectrum in the BB sector.
The main technical difficulty in treating the finite-$\omega$ case
is that any nonzero $\omega$ renders the continuous BB spectrum
be discrete (assuming $\Omega$ positive).
This is related to the noncompactness of the BB sector
eventually responsible for localization.

The zero mode of the Hamiltonian (\ref{H}) solving $H\Psi_0=0$
has been recently obtained in a closed form valid
for all frequencies:\cite{SO07}
\be
\label{Psi0}
  \Psi_0(\lambda_F,\lambda_B)
  =
  I_0(q) p K_1(p) + q I_1(q) K_0(p) ,
\ee
where
\be
\label{pq}
  p = \sqrt{8\Omega(\lambda_B+1)} ,
\qquad
  q = \sqrt{8\Omega(\lambda_F+1)} .
\ee

The function $A(\omega,t)$ which determines the large-distance
behavior of the LDOS correlation function (\ref{R-A})
can be conveniently written in terms of
the expectation value of a certain evolution operator\cite{Efetov-book,DM}
calculated with respect to the zero mode (\ref{Psi0}):
\be
\label{A}
  A(\omega,t) = \frac12 \Re \corr{\Psi_0|e^{-2\tilde Ht}|\Psi_0} ,
\ee
where $\corr{\cdot|\cdot}$ is the flat bilinear form on singlet states:
\be
  \corr{\phi|\chi}
  =
  \int_{-1}^1 d\lambda_F \int_0^\infty d\lambda_B \,
  \phi(\lambda_B,\lambda_F)
  \chi(\lambda_B,\lambda_F)
  .
\ee
The Hamiltonian
$\tilde H = (\lambda_B-\lambda_F)^{-1} H (\lambda_B-\lambda_F)$
splits into the sum of the bosonic and fermionic parts:
\be
\label{tilde-H}
  \tilde H = \tilde H_B + \tilde H_F ,
\ee
where
\begin{align}
\label{tilde-HB}
  & \tilde H_B
  = - \frac12 \partial_{\lambda_B} (\lambda_B^2-1) \partial_{\lambda_B}
  + \Omega \lambda_B
  ,
\\
\label{tilde-HF}
  & \tilde H_F
  = - \frac12 \partial_{\lambda_F} (1-\lambda_F^2) \partial_{\lambda_F}
  - \Omega \lambda_F .
\end{align}

Analytic calculation of the expectation value in (\ref{A}) as a function of
arbitrary $\omega$ and $t$ is a complicated task.
The main technical problem is that the eigenfunctions of the Hamiltonians
$\tilde H_B$ and $\tilde H_F$ (known in the theory as a particular class
of the Coulomb spheroidal functions\cite{KPS76}) cannot be found explicitly
for an arbitrary value of the parameter $\Omega$.

In the limit $\Omega\gg1$, only small deviations of $\lambda_F$ and
$\lambda_B$ from 1 are important, and the correlation function
$R(\omega,t)$ can be calculated perturbatively in $\Delta_\xi/\omega$.
In this regime the LDOS correlations can be found within
the standard diagrammatic technique by expanding in the diffusive
modes \cite{AltshulerShklovskii}.

In the most interesting case $\Omega\ll1$,
a naive application of the perturbation theory in $\Omega$
fails in the ``bosonic'' sector.
Smallness of $\Omega$ in the last term in (\ref{tilde-HB})
is compensated by large values of $\lambda_B$,
and thus the term $\Omega\lambda_B$ in $\tilde H_B$
cannot be considered as a small perturbation.
An accurate construction of the perturbation theory in
the bosonic sector at $\Omega\ll1$ is a subject of Sec.~\ref{S:Bosonic}.

Though physically relevant $\Omega$ is imaginary,
see Eq.~(\ref{Omega-def}),
we will perform calculations assuming that $\Omega$
is a real positive number
and make the analytic continuation to imaginary $\Omega$
at the final stage of the calculation.
With real $\Omega$, both $\tilde H_B$ and $\tilde H_F$ become Hermitian
operators with a discrete spectrum,
that allows to write the spectral decomposition
of the expectation value (\ref{A}):
\be
\label{A-spectral}
  A(\omega,t)
  =
  \frac12
  \Re \sum_{mk}
  \frac{\corr{\Psi_0|\chi_m\phi_k}^2}
    {\corr{\chi_m|\chi_m} \corr{\phi_k|\phi_k}}
  e^{-2(E_m+E_k)t} ,
\ee
where $\chi_m(\lambda_F)$ and $\phi_k(\lambda_B)$ are the eigenfunctions
of the Hamiltonians $\tilde H_F$ and $\tilde H_B$:
\begin{align}
  \tilde H_F \phi_m(\lambda_F) & = E_m \phi_m(\lambda_F) ,
\\
  \tilde H_B \phi_k(\lambda_B) & = E_k \phi_k(\lambda_B) .
\end{align}

In the next sections we perturbatively construct
the eigenfunctions of the Hamiltonians
(\ref{tilde-HB}) and (\ref{tilde-HF})
and calculate the matrix elements
\begin{multline}
\label{overlap-with-Psi0}
  \corr{\Psi_0|\chi_m\phi_k}
  =
  \corr{I_0(q)|\chi_m}
  \corr{pK_1(p)|\phi_k}
\\
  +
  \corr{qI_1(q)|\chi_m}
  \corr{K_0(p)|\phi_k}
\end{multline}
needed to evaluate (\ref{A-spectral}).

In the fermionic sector, the perturbation theory is straightforward,
while its application in the bosonic sector is much more involved
due to its non-compactness. Nevertheless we start with the bosonic
sector in Sec.~\ref{S:Bosonic} and use some notations introduced
there in dealing with the fermionic sector in Sec.~\ref{S:Fermionic}.

\section{Eigensystem of the ``bosonic'' Hamiltonian $\tilde H_B$
at small $\Omega\ll1$}
\label{S:Bosonic}

The main technical problem in studying the noncompact bosonic sector
is that it cannot be decomposed into
a ``regular part'' and a perturbation
such that (i) the regular part is exactly solvable
and (ii) the perturbation is small on the whole interval $1\leq\lambda<\infty$
(in this section we denote the bosonic variable $\lambda_B$ by $\lambda$
for brevity).
Therefore to have a well-controlled perturbation theory
one should use different decompositions
of the Hamiltonian (\ref{tilde-HB}) at small and large $\lambda$
in order to capture the behavior of the wave function
near the singular points $\lambda=1$ and $\lambda=\infty$.

At small $\lambda \ll 1/\Omega$, we will decompose the Hamiltonian
as $\tilde H_B = \tilde H_{B1}^0 + V_{B1}$ with
\be
  \tilde H_{B1}^0
  = - \frac12 \partial_{\lambda} (\lambda^2-1) \partial_{\lambda} ,
\qquad
  V_{B1} = \Omega \lambda ,
\ee
whereas at large $\lambda \gg 1$ we adopt the decomposition
$\tilde H_B = \tilde H_{B2}^0 + V_{B2}$ with
\begin{multline}
  \tilde H_{B2}^0
  = - \frac12 \partial_{\lambda} (\lambda+1)^2 \partial_{\lambda} + \Omega(\lambda+1),
\\
  V_{B2} = \partial_{\lambda} (\lambda+1) \partial_{\lambda} - \Omega .
\end{multline}

\subsection{Small $\lambda$: Legendre scattering states}

The Hamiltonian $\tilde H_{B1}^0$ has a continuous spectrum
\be
\label{E0}
  E^{(0)}_k = \frac{k^2}{2} + \frac18 ,
\ee
($k$ is a real number) with the eigenfunctions finite at $\lambda=1$ given by
\be
  L_k^{(0)}(\lambda) = P_{-1/2+ik}(\lambda) ,
\ee
where $P_{-1/2+ik}(\lambda)$ is the Legendre function
(with such an index it is usually referred to as the conical function).
The standard perturbation theory with respect to the perturbation $V_{B1}$
is ill-defined since the matrix elements of the perturbation diverge
when integrated over the whole semiaxis $\lambda>1$.
Nevertheless it is possible to construct the wave function
perturbatively in the limit $1<\lambda\ll1/\Omega$.
Using the properties of Legendre functions, one can show
that $V_{B1}$ acting on the bare function $L_k^{(0)}(\lambda)$
shifts its ``momentum'' $k$ by $\pm i$:
\be
\label{VL}
  V_{B1} L_k^{(0)}(\lambda)
  =
  \Omega
  \frac{(ik+\frac12)L_{k-i}^{(0)}(\lambda)+(ik-\frac12) L_{k+i}^{(0)}(\lambda)}{2ik}
  .
\ee
Therefore we can search the exact solution as a formal series
\be
\label{L-series}
  L_k(\lambda) = \sum_{n=-\infty}^\infty c_n(\Omega,k) L_{k+in}^{(0)}(\lambda) .
\ee
Then acting by the Hamiltonian $\tilde H_B$
and using that $L_{k+in}^{(0)}(\lambda)$
is also an eigenfunction of $\tilde H_{B1}^0$
we obtain a tridiagonal system of linear equations
for the coefficients $c_n(\Omega,k)$:
\begin{multline}
\label{hc}
  \left[ \frac{(k+in)^2}{2} + \frac18 - E_k \right] c_n
\\
+ \frac{\Omega(ik-n+\frac12)}{2(ik-n+1)} c_{n-1}
+ \frac{\Omega(ik-n-\frac12)}{2(ik-n-1)} c_{n+1}
= 0
..
\end{multline}
The solution (\ref{L-series}) can be multiplied by an arbitrary
$\lambda$-independent constant. To fix this freedom, we assume
\be
  c_0(\Omega,k)=1
\label{normalization-convention}
\ee
and will normalize the eigenfunction $\phi_k(\lambda)$ later on,
see Sec.~\ref{SS:normalization}.

The system of equations (\ref{hc}) allows to obtain the series
expansion of $c_n(\Omega,k)$ in powers of $\Omega$, starting
with $\Omega^{|n|}$. In particular,
\begin{align}
  &c_1(\Omega,k)
  =
    \frac{i\Omega}{2k}
  - \frac{i(2k^2+7ik-7) \Omega^3}{32(k-i)k(k+i)^2(k+2i)}
  +
  \dots
  ,
\\
  &c_2(\Omega,k)
  =
  - \frac{(2k+3i) \Omega^2}{16 k (k+i)^2}
  +
  \dots
  ,
\\
  &c_3(\Omega,k)
  =
  - \frac{i(2k+5i) \Omega^3}{96 k(k+i)^2(k+2i)}
  +
  \dots
\end{align}
The coefficients with negative indices can be found from
\be
  c_{-n}(\Omega,k)
  =
  c_{n}(\Omega,-k) .
\label{c-n=}
\ee

The energy $E_k$ also acquires corrections in powers of $\Omega$.
Using Eq.~(\ref{hc}), it can be expressed in terms of $c_{\pm1}(\Omega,k)$:
\be
\label{Ek-via-c}
  E_k
  =
  E_k^{(0)}
+ \frac{\Omega(ik+\frac12)}{2(ik+1)} c_{-1}(\Omega,k)
+ \frac{\Omega(ik-\frac12)}{2(ik-1)} c_{1}(\Omega,k)
..
\ee
Thus $E_k$ is given by the series in $\Omega^2$:
\be
\label{Ek}
  E_k
  =
  E_k^{(0)}
  + \frac{\Omega^2}{4(k^2+1)}
  + \frac{(5k^2+7)\Omega^4}{44(k^2+1)^3(k^2+4)}
  + \dots
\ee

\subsection{Large $\lambda$: Bessel scattering states}

The eigenfunctions of the Hamiltonian $\tilde H_{B2}^0$
which decay at $\lambda\to\infty$ are expressed in terms
of the modified Bessel function of the second kind
(MacDonald function):
\be
  B_k^{(0)}(\lambda) = \frac{K_{2ik}(p)}{p}
\ee
where $p=\sqrt{8\Omega(\lambda+1)}$.
The corresponding energy is given by Eq.~(\ref{E0}).
In order to take the perturbation $V_{B2}$ into account
we note that its action on $B_k^{(0)}(\lambda)$
is equivalent to shifting the index $k$ by $\pm i$:
\be
\label{VB}
  V_{B2} B_k^{(0)}(\lambda)
  =
  \Omega \frac{(ik+\frac12)B_{k-i}^{(0)}(\lambda)+(ik-\frac12) B_{k+i}^{(0)}(\lambda)}{2ik}
\ee
[Note the spectacular coincidence with Eq.~(\ref{VL})!]
Therefore the solution of $\tilde H_B B_k(\lambda)=E_k B_k(\lambda)$
can be naturally written as a series
\be
\label{B-series}
  B_k(\lambda) = \sum_{n=-\infty}^\infty c_n(\Omega,k) B_{k+in}^{(0)}(\lambda)
\ee
where the coefficients $c_n(\Omega,k)$ coincide exactly with the
coefficients of expansion over Legendre functions in (\ref{L-series}),
and can be found from the set of linear equations (\ref{hc}).
The spectrum $E_k$ as a series in $\Omega$ can be found from
the same equation, see Eq.~(\ref{Ek}).

\subsection{Wave functions of the discrete spectrum}

The Legendre and Bessel scattering states constructed above
provide the solutions to the Hamiltonian (\ref{tilde-HB})
with the proper behavior at $\lambda\to1$ and $\lambda\to\infty$, respectively.
These solutions exist for arbitrary ``wave vectors'' $k$.
The wave function $\phi_k(\lambda)$ satisfying both boundary conditions
can be found only for a discrete set of $k$.
Such a wave function can be conveniently represented as
\be
  \phi_k(\lambda)
  =
  \begin{cases}
    L_k(\lambda) , & \lambda<\lambda_* , \\
    C_k B_k(\lambda) , & \lambda>\lambda_* ,
  \end{cases}
\label{phik}
\ee
with some arbitrary $\lambda_*$ and a suitably chosen factor $C_k$.
Matching the left- and right-scattering states,
$L_k(\lambda)$ and $C_kB_k(\lambda)$,
at $\lambda_*$ provides the condition for the allowed values of $k$.

In the limit $\Omega\ll1$, determination of the discrete spectrum
is simplified by the fact that there exists a broad intermediate
region $1 \ll \lambda \ll 1/\Omega$,
where both $L_k(\lambda)$ and $B_k(\lambda)$
can be represented by their asymptotic expressions.
To find them we write each term in Eqs.~(\ref{L-series})
and (\ref{B-series}) as a superposition of right- and left-moving waves:
\begin{align}
 L^{(0)}_k(\lambda&)
  = L^{(0)+}_k + L^{(0)-}_k, \\
 B^{(0)}_k(\lambda&)
  = B^{(0)+}_k + B^{(0)-}_k,
\end{align}
and represent the right-moving waves
$L^{(0)+}_k$ and $B^{(0)+}_k$
by the following hypergeometric series:
\begin{gather}
 L^{(0)+}_k
  = \sum_{m=0}^\infty \frac{(-1)^m \Gamma(2ik - m)}
    {m!\, \Gamma^2 \big( \frac{1}{2} + ik - m \big)} \left(
      \frac{\lambda + 1}{2}
    \right)^{-\frac{1}{2} + ik - m},
\label{L-hypergeometric}
\\
 B^{(0)+}_k
  = \sum_{m=0}^\infty \frac{(-1)^m}{4 m!} \Gamma(-2ik - m) \big[
      2\Omega (\lambda + 1)
    \big]^{-\frac{1}{2} + ik + m} .
\label{B-hypergeometric}
\end{gather}
The left-moving waves are expressed by the right-moving waves as
$L^{(0)-}_k = L^{(0)+}_{-k}$ and $B^{(0)-}_k = B^{(0)+}_{-k}$.

The right-moving waves in the Legendre and Bessel scattering states
should be matched by a suitably chosen factor:
\be
  L_k^{+}(\lambda) = C^+_k B_k^{+}(\lambda) .
\ee
We compute the value of $C^+_k$ in the following manner. First we substitute the
asymptotic expansion Eq.~(\ref{L-hypergeometric}) of the Legendre function into
Eq.\ (\ref{L-series}). This yields a double series with the terms $\sim (\lambda
+ 1)^{-1/2 + ik - m - n}$. Substituting the Bessel asymptotics
(\ref{B-hypergeometric}) into the expansion (\ref{B-series}) yields another
double series with the terms $\sim (\lambda + 1)^{-1/2 + ik + m - n}$. In order
to match these two asymptotic expansions, we pick the terms with $n = -m$ from
the Legendre series and with $n = m$ from Bessel series. This way we obtain two
different representations of the coefficient multiplying $(\lambda + 1)^{-1/2 +
ik}$ in the common asymptotic region $1 \ll \lambda \ll 1/\Omega$. They are
single series in $c_n$, that is in powers of $\Omega$. By equating the two
expressions we find $C^+_k$. The other terms in the asymptotic expansion of
$L_k^+$ and $B_k^+$ will automatically match due to the properties of the
coefficients $c_n(\Omega,k)$. Thus we have
\be
  C^+_k
  =
  \frac{4 \Gamma(2 ik) (4\Omega)^{1/2 - ik}}
    {\Gamma(-2ik)\Gamma^2(1/2 + ik)}
  e^{i\Phi(k)} ,
\label{Ck-right-matching}
\ee
where the phase factor
\be
  e^{i\Phi(k)}
  =
  \frac{\displaystyle
    \sum_{n=0}^\infty c_n(\Omega,-k) \frac{(-1)^n}{n!}
    \frac{\Gamma(2ik + n)}{\Gamma(2ik)}
  }{\displaystyle
    \sum_{n=0}^\infty c_n(\Omega,k) \frac{(-1)^n}{n!}
    \frac{\Gamma(-2ik + n)}{\Gamma(-2ik)}
  }
\label{phase-factor}
\ee
has a regular series expansion in powers of $\Omega^2$:
\be
  e^{i\Phi(k)}
  =
  1
  +
  \frac{ik\,\Omega^2}{4(k^2+1)^2}
  +
  \dots
\ee

The waves in the opposite direction, $L^-_k(\lambda)$ and $B^-_k(\lambda)$,
are matched by the coefficient
\begin{equation}
 C^-_k
  = C^+_{-k} .
\end{equation}
The full function $\phi_k$ can be matched only if the two coefficients
coincide, $C^+_k = C^-_k$, which is a condition for the discrete spectrum.
It can be conveniently formulated as the condition $S(k)=1$ for the
scattering matrix $S(k) \equiv C^+_k/C^-_k$:
\be
\label{S(k)}
  S(k)
  =
  S_0(k)
  e^{2i\Phi(k)} ,
\ee
where
\be
\label{S0(k)}
  S_0(k)
  =
  \left(4\Omega \right)^{-2ik} \left[
    \frac{\Gamma(2ik)\Gamma(1/2-ik)}{\Gamma(-2ik)\Gamma(1/2+ik)}
  \right]^2 .
\ee

Equation $S(k)=1$ determines the values of the wave vector $k$
corresponding to the states of the discrete spectrum.
For the low-lying states ($k\lesssim1$), one finds
\be
\label{kn}
  k \simeq \frac{\pi N_k}{\ln(1/4\Omega)} ,
\ee
with $N_k=1,2,\dots$ labelling the energy level
($N_k=1$ refers to the ground state).
According to Eq.~(\ref{kn}), the lowest discrete $k$'s are nearly
equidistant. This can be
naturally seen if we rewrite the Hamiltonian $\tilde H_B$
in terms of the angular variable $\theta=\mathop{\rm arccosh}\lambda$.
The motion in the resulting potential will resemble the motion
in a box of size $\ln(1/4\Omega)$, leading to the quantization rule (\ref{kn}).

At the points of the discrete spectrum,
the two coefficients $C^+_k$ and $C^-_k$ coincide,
and we denote them simply by $C_k$. A calculation of
$C_k^2=C^+_k C^-_k$ using Eq.~(\ref{Ck-right-matching}) gives
\be
\label{C_k}
  C_k
  =
  \sigma_k
  \frac{8\sqrt{\Omega}}{\pi}
  \cosh\pi k ,
\qquad
  \sigma_k=(-1)^{N_k} ,
\ee
where $N_k$ is the number of the discrete energy level, see Eq.~(\ref{kn}).

Remarkably, the expression (\ref{C_k}) does not have any corrections in $\Omega$
and is exact.

\subsection{Normalization of $\phi_k(\lambda)$}
\label{SS:normalization}

The eigenfunction $\phi_k(\lambda)$ perturbatively constructed above
does not have the unit norm $\corr{\phi_k|\phi_k}$.
Instead, by definition, it is normalized in such a way that
its expansion (\ref{L-series}) in Legendre functions comes
with $c_0(\Omega,k)=1$. Therefore we further need to compute
the norm of $\phi_k(\lambda)$. This can be done by writing
the function $\phi_k(\lambda)$ in the form (\ref{phik})
and calculating the normalization integrals with
$[L_k(\lambda)]^2$ and $[B_k(\lambda)]^2$.
This program is performed in Appendix \ref{A:normalization}
leading to a surprisingly simple expression:
\be
  \langle \phi_k | \phi_k \rangle
  =
  -i\frac{\coth \pi k}{2\pi}
  \frac{\partial \ln S(k)}{\partial k} \sum_{n=-\infty}^\infty
  \frac{[c_n(\Omega,k)]^2}{k + i n} ,
\label{<phi|phi>}
\ee
where $S(k)$ is the exact scattering matrix (\ref{S(k)}).

(Strictly speaking, Eq.~(\ref{<phi|phi>}) has not been proven.
It has been verified to the order $O(\Omega^6)$,
see Appendix \ref{A:normalization} for discussion.
Nevertheless, relying on the anticipated duality between the bosonic
and fermionic sectors of a supersymmetric theory,
we believe that Eq.~(\ref{<phi|phi>}) is exact.)

\subsection{Overlap with the zero mode}

The last ingredient required for the evaluation of the general
expression (\ref{A-spectral}) is the overlap integral $\corr{\Psi_0|\phi_k}$
with the zero mode $\Psi_0$.
According to Eq.~(\ref{overlap-with-Psi0}), this involves
two matrix elements from the bosonic sector:
$\corr{K_0(p)|\phi_k}$ and $\corr{pK_1(p)|\phi_k}$.
They are calculated in Appendix \ref{A:overlap}, and the result
has the form:
\begin{align}
\label{K0phik}
  \corr{K_0(p)|\phi_k}
  & =
  \sigma_k \pi
  \sum_{n = -\infty}^\infty
  \frac{c_n(\Omega,k)}{2\sqrt{\Omega} \cosh \pi k} ,
\\
\label{K1phik}
  \corr{pK_1(p)|\phi_k}
  & =
  \sigma_k \pi
  \sum_{n = -\infty}^\infty
  \frac{c_n(\Omega,k) [1+4(k+in)^2]}{4\sqrt{\Omega} \cosh \pi k} .
\end{align}
The overall sign factor $\sigma_k=\pm1$ defined in Eq.~(\ref{C_k})
will drop from the final expression (\ref{A-spectral}).

\section{Eigensystem of the ``fermionic'' Hamiltonian $\tilde H_F$
at small $\Omega\ll1$}
\label{S:Fermionic}

In the fermionic sector, the standard perturbation theory
in small $\Omega$ can be easily developed by treating
the $\Omega$-dependent term in the Hamiltonian (\ref{tilde-HF})
as a perturbation. However, instead of doing these routine calculations,
one can notice a formal duality between the bosonic and fermionic sectors.
By substituting $\lambda_B\mapsto\lambda_F$,
one finds $\tilde H_B\mapsto-\tilde H_F$.
Therefore, the fermionic eigenfunctions $\chi_m(\lambda_F)$
can be readily obtained from the Legendre-series expansion
(\ref{L-series}) in the bosonic sector by imposing the condition
of regularity at $\lambda_F=-1$. The latter condition
restricts $k$ to $-i(m+1/2)$ with $m=0,1,\dots$,
rendering Legendre functions $P_{-1/2+ik}(\lambda_F)$
be Legendre polynomials $P_m(\lambda_F)$.

Thus the fermionic eigenfunctions can be written with the help
of the coefficients $c_n(\Omega,k)$ as
\be
\label{chi-series-c}
  \chi_m(\lambda_F)
  =
  \sum_{n=-\infty}^\infty c_n(\Omega,i(m+1/2)) P_{m+n}(\lambda_F) ,
\ee
where we use the symmetry relation (\ref{c-n=}).
Since $P_{-l-1}(x)=P_l(x)$, we can rewrite Eq.~(\ref{chi-series-c})
as a sum of Legendre polynomials with nonnegative indices:
\be
\label{chi-series}
  \chi_m(\lambda_F)
  =
  \sum_{l=0}^\infty \gamma_l(\Omega,m) P_{l}(\lambda_F) ,
\ee
where the coefficients $\gamma_l(\Omega,m)$ are defined as
\be
  \gamma_l(\Omega,m)
  =
    c_{l-m}(\Omega,i(m+1/2))
  + c_{-1-l-m}(\Omega,i(m+1/2))
  .
\label{gamma-def}
\ee
The eigenenergy $E_m$ is given by $-E_k$ [see Eq.~(\ref{Ek-via-c})]
with the substitution $ik\mapsto m+1/2$:
\begin{multline}
\label{Em-via-c}
  E_m
  =
  \frac{m(m+1)}{2}
- \frac{\Omega(m+1)}{2m+3} c_{1}(\Omega,i(m+1/2))
\\ {}
- \frac{\Omega m}{2m-1} c_{-1}(\Omega,i(m+1/2))
..
\end{multline}

The wave function (\ref{chi-series}) is normalized according to
\be
\label{<chim|chim>}
  \corr{\chi_m|\chi_m}
  =
  \sum_{l=0}^\infty
  \frac{2}{2l+1}
  [\gamma_l(\Omega,m)]^2 .
\ee

The overlap integrals of $\chi_m(\lambda_F)$ with $I_0(q)$ and $qI_1(q)$
can be calculated with the help of the expansion (\ref{chi-series})
and the following exact expressions:
\begin{subequations}
\label{<I|P>}
\begin{align}
  \corr{I_0(q)|P_l(\lambda_F)}
  & = \sqrt{\frac{2l+1}{2}} \frac{I_{2l+1}(4\sqrt\Omega)}{\sqrt\Omega} ,
\\
  \corr{qI_1(q)|P_l(\lambda_F)}
  & = \sqrt{\frac{2l+1}{2}} \Omega \frac{\partial}{\partial\Omega}
  \frac{2I_{2l+1}(4\sqrt\Omega)}{\sqrt\Omega} .
\end{align}
\end{subequations}

~

\section{Behavior of $R(\omega,t)$ at small $\omega$ and at arbitrary $t$}
\label{S:result}

\subsection{General expression}

Now we are in a position to calculate $A(t)$ with the help
of the spectral decomposition (\ref{A-spectral}).
Since we are aimed at making the analytic continuation
to imaginary $\Omega$, it is desirable to get rid of the sum
over discrete spectrum in the bosonic sector existing
only for real $\Omega$. This can be done with the help
of the identity
\be
  \sum_k \cdots
  =
  -i \sum_{n=-\infty}^\infty
  \int_0^\infty \frac{dk}{2\pi} \,
  S^n(k) \frac{\partial \ln S(k)}{\partial k}
  \cdots ,
\label{sum2int}
\ee
where we utilized the fact that the scattering matrix $S(k)=1$
at the points of discrete spectrum and
employed the Poisson resummation formula.

As a result, the factor $\partial\ln S(k)/\partial k$
appearing in Eq.~(\ref{sum2int}) cancels $\partial\ln S(k)/\partial k$
in the normalization (\ref{<phi|phi>}),
and the general expression (\ref{A-spectral}) can be written
in the following form:
\begin{widetext}
\begin{subequations}
\label{A-M}
\begin{gather}
  A(\omega,t)
  =
  \sum_{n=0}^\infty
  A^{(n)}(\omega,t) ,
\\
\label{A0}
  A^{(0)}(\omega,t)
  =
  4\pi^2
  \Re
  \sum_{m=0}^\infty
  \int_0^\infty dk \, k \, \frac{\tanh\pi k}{\cosh^2\pi k}
  M_{mk}(\Omega) e^{-2(E_k+E_m)t} ,
\\
  A^{(n>0)}(\omega,t)
  =
  4\pi^2
  \Re
  \sum_{m=0}^\infty
  \int_{-\infty}^\infty dk \, k \, \frac{\tanh\pi k}{\cosh^2\pi k} S^n(k)
  M_{mk}(\Omega) e^{-2(E_k+E_m)t} .
\label{An>0}
\end{gather}
\end{subequations}
In deriving (\ref{An>0}) we used the symmetry of the scattering matrix,
$S^{-n}(k)=S^n(-k)$, which allowed us to extend the integral to the whole
real axis.

In Eqs.~(\ref{A-M}), the scattering matrix $S(k)$
is defined in Eq.~(\ref{S(k)}),
the energies $E_k$ and $E_m$ are given by Eqs.~(\ref{Ek-via-c})
and (\ref{Em-via-c}),
and the function $M_{mk}(\Omega,k)$ coming from the matrix elements
in Eq.~(\ref{A-spectral}) has a regular expansion in powers of
$\Omega$ which can be calculated with the help of
Eqs.~(\ref{<phi|phi>}), (\ref{K0phik}), (\ref{K1phik}),
(\ref{chi-series}), (\ref{<chim|chim>}) and (\ref{<I|P>}):
\be
\label{Mmk}
  M_{mk}(\Omega)
  =
  \frac{
    \displaystyle
    \left(
    \sum_{l=0}^\infty
    \sum_{s=-\infty}^\infty
    \frac{\gamma_{l}(\Omega,m) c_{s}(\Omega,k)}{2l+1}
    \left\{
      \left[
        l+\frac14+(k+is)^2
      \right]
      I_{2l}(4\sqrt{\Omega})
    +
      \left[
        l+\frac34-(k+is)^2
      \right]
      I_{2l+2}(4\sqrt{\Omega})
    \right\}
    \right)^2
  }
  {
    \displaystyle
    4 \Omega
    \left(
      \sum_{l=0}^\infty
      \frac{1}{2l+1} \gamma_{l}^2(\Omega,m)
    \right)
    \left(
      \sum_{s=-\infty}^\infty
      \frac{k}{k+is} c_{s}^2(\Omega,k)
    \right)
  }
  .
\ee
\end{widetext}
Equations (\ref{A-M}) provide a general expression for the function
$A(\omega,t)$ in the limit of small $\omega$.

The term $A^{(0)}(\omega,t)$ given by Eq.~(\ref{A0})
is an even function of $\Omega$ responsible for
the decay of the LDOS correlations at the localization
length scale, see Eq.~(\ref{A00}) below.

The terms $A^{(n)}(\omega,t)$ with $n>0$ are given by the
integrals (\ref{An>0}) with the strongly oscillating function
$S_0^n(k) \propto S_0^n(k)\propto\Omega^{-2ink}$.
Therefore they can be naturally calculated by deforming
the integration contour to reach the saddle point
\be
  k_*^{(n)} = \frac{i n \ln(1/4\Omega)}{t} ,
\label{k*}
\ee
In the process of contour deformation one would pick up the
contribution of the poles of the integrand in (\ref{An>0})
located at $k=i/2$, $i$, $3i/2,\dots$
The interplay between the contributions of the saddle point
and of the poles is responsible for the behavior of
the LDOS correlator at the Mott scale $L_M$
given by Eq.~(\ref{L-Mott}).

In Figure \ref{F:Mott}, we plot the spatial dependence of the LDOS
correlation function $R(\omega,x) = 1+A(\omega,t)$ calculated numerically
with the help of Eq.~(\ref{A-M}), where it is crucial to take the real
part and use the fact that $\Omega$ is purely imaginary according to
Eq.~(\ref{Omega-def}).
Different curves on the graph correspond to various
values of $\omega$ in the deeply localized region, $\omega \ll \Delta_\xi$.
The smaller is the ratio $\omega/\Delta_\xi$, the faster do the sums
in Eq.~(\ref{A-M}) converge.

The qualitative behavior of the LDOS correlation
function in the quasi-1D geometry coincides with the behavior
in the strict 1D geometry \cite{GDP83,SI87}:
On increasing $x$, the function $R(\omega,x)$ first decays
at the localization length and then
reaches the uncorrelated value $R=1$ at the Mott scale
$L_M$.

\subsection{Leading order in $\omega$}

In this section we evaluate Eq.~(\ref{A-M}) analytically
in the limit of vanishing frequency, $\omega/\Delta_\xi\to0$.
Then the localization length and the Mott scale are well
separated and one can obtain simple expressions for $R(\omega,x)$
describing its decay at $x\sim\xi$
and further growth near $x\sim L_M$.

It can be easily seen that the coefficients $\gamma_l(\Omega,m)$
decrease with increasing of the fermionic ``momentum'' $m$:
$M_{mk}(\Omega) = O(\Omega^{2m-1})$.
Therefore, only the ground state ($m=0$) of the fermionic sector
contributes to $A(\omega,t)$ in the limit $\omega\to0$:
\be
\label{M0}
  M_{0k}(\Omega)
  =
  \left(
    \frac{1}{4\Omega} + 1
  \right)
  (k^2+1/4)^2
  +
  O(\Omega)
  .
\ee
To the leading order in small $\omega$ one can also
replace $S(k)\to S_0(k)$, $E_k\to E_k^{(0)}$, $E_m\to E_m^{(0)}$,
since corrections to these quantities start with $\Omega^2$.

Consider first the contribution of $A^{(0)}(\omega,t)$.
Since $\Re\Omega=0$, the ``dangerous'' term $1/4\Omega$
in (\ref{M0}) does not contribute and we may safely
put $\omega=0$:
\be
\label{A00}
  A^{(0)}(0,t)
  =
  4\pi^2 \frac{\partial^2}{\partial t^2}
  \int_0^\infty k \, dk \, \frac{\tanh\pi k}{\cosh^2\pi k}
  e^{-(k^2+1/4)t}
  .
\ee

The terms $A^{(n)}(\omega,t)$ with $n>0$ can be evaluated by deforming
the integration contour to pass through the saddle point (\ref{k*})
and picking up the contributions of poles to be crossed.
Since the residue at the pole $ip/2$ is proportional to $\Omega^{np-1}$,
only the term with $n=1$ is relevant to the leading order in $\omega$:
\begin{multline}
\label{A-11}
  A^{(1)}(\omega,t)
  =
  4\pi^2
  \Re
  \int_{-\infty}^\infty dk \, k \, \frac{\tanh\pi k}{\cosh^2\pi k}
  S_0(k)
  \left(
    \frac{1}{4\Omega} + 1
  \right)
\\ {}
  \times
  (k^2+1/4)^2 e^{-(k^2+1/4)t}
  +
  O(\Omega).
\end{multline}

The value of the integral (\ref{A-11}) depends on the relative
position between the saddle point $k_*^{(1)}=i(t_M+i\pi)/2t$
and the pole $i/2$. They nearly merge at $t\sim t_M$, where
\be
  t_M
  =
  2 \Re \ln \frac{1}{4\Omega}
  =
  2\ln\frac{\Delta_\xi}{\omega}
  \gg 1
\ee
corresponds to the Mott scale $L_M$ given by Eq.~(\ref{L-Mott}).
Thus, within the present technique, the appearance of the Mott scale
is related to competition between the pole and the saddle point.

The contribution of the saddle point $k_*^{(1)}$ calculated
with exponential accuracy is
\be
\label{A1-saddle}
  A^{(1)}(\omega,t) \Big|_\text{s.~p.~$k_*^{(1)}$}
  \sim
  \exp \left( -\frac{(t - t_M)^2}{4t} \right).
\ee
The contribution of the pole $i/2$ to $A^{(1)}(\omega,t)$
remains finite in the limit $\omega\to0$:
\be
\label{A-11small}
  A^{(1)}(0,t) \Big|_\text{pole at $i/2$}
  =
  -1
  .
\ee
The pole contribution (\ref{A-11small}) should be taken
into account only for $t<t_M$, since at larger $t$,
$\Im k_*^{(1)}<1/2$ and the pole $i/2$ should not be
crossed in deforming the integration contour to the saddle point.

Various asymptotic regions for the correlator $R(\omega,x)$
are considered below.

\subsubsection{Decay of correlations at $t<t_M/2$}

At spatial separations $x$ smaller than the Mott scale,
$t<t_M$, the LDOS correlation function (in the leading
order in $\omega$) is determined by
the interplay of $A^{(0)}(0,t)$ given by Eq.~(\ref{A00}),
and the two contributions to $A^{(1)}(\omega,t)$
[Eq.~(\ref{A1-saddle}) and (\ref{A-11small})].
The pole contribution (\ref{A-11small}) cancels 1 in Eq.~(\ref{R-A}).
Therefore, $R(\omega,x)$ is determined by the competition
between Eqs.~(\ref{A00}) and (\ref{A1-saddle}).
At $t<t_M/2$, the former dominates and the LDOS correlation
function reduces to $A^{(0)}(0,t)$ given by Eq.~(\ref{A00}):
\be
\label{R-decay}
  R(0,x)
  =
  4\pi^2 \frac{\partial^2}{\partial t^2}
  \int_0^\infty k \, dk \, \frac{\tanh\pi k}{\cosh^2\pi k}
  e^{-(k^2+1/4)t}
  .
\ee
This function decays at the localization length scale
and describes the limiting curve in Fig.~(\ref{F:Mott})
as $\omega\to0$.

The formula (\ref{R-decay}) is well known in the theory of localization
in one-dimensional systems.
It describes the decay of a single wave function, both
in the 1D \cite{Gogolin,LGP82,Kolokolov1995}
and Q1D \cite{EL83,Mirlin-JMP-1997,Efetov-book} geometries.
In the two limiting cases of small and large $t$ one finds:
\be
\label{R0-asymp}
 R(0,x)
  = \begin{cases}
     \dfrac{2}{3} - \dfrac{2t}{3} + \dfrac{8t^2}{15} + \dots, & t \ll 1, \\[9pt]
     \dfrac{\pi^{7/2}}{16 t^{3/2}}\; e^{-t/4}, & t \gg 1.
    \end{cases}
\ee

\subsubsection{Intermediate asymptotics at $t_M/2<t<t_M$}

The LDOS correlator decays according to Eq.~(\ref{R-decay})
only for $t<t_M/2$. For larger values of $t$, the saddle point
contribution (\ref{A1-saddle}) should be taken into account
on the background of the exponentially small $A^{(0)}(0,t)$.
They become comparable at $t\approx t_M/2$, which results
in a minimum on the $R(\omega,x)$ curve, of the order of
$e^{-t_M/8} \sim (\omega/\Delta_\xi)^{1/4}$.
At larger distances ($t_M/2<t<t_M$),
the LDOS correlator increases as
\be
\label{R-intermed}
 R(\omega, x)
  \sim \exp \left( -\frac{(t - t_M)^2}{4t} \right).
\ee

\subsubsection{Behavior at the Mott scale, $t\approx t_M$}

In the vicinity of the Mott scale, at $|t-t_M|\sim \sqrt{t_M}$,
the saddle point $k_*^{(1)}$ approaches $i/2$, and it becomes
impossible to treat the pole at $k=i/2$ independently of the saddle point.
To evaluate (\ref{A-11}) in this region we note that its first
derivative over $t$ does not have a pole at $k=i/2$ and thus can
be easily found with the steepest descent method:
\be
\label{dA/dt}
  \frac{\partial A^{(1)}(\omega,t)}{\partial t}
  =
  \frac{1}{2\sqrt{\pi t_M}}
  \exp\left(
    - \frac{(t-t_M)^2}{4t_M}
  \right) .
\ee
Integrating (\ref{dA/dt}) in the vicinity of $t_M$,
we get
\be
\label{R-Mott}
  R(\omega,x)
  =
  \frac12
  +
  \frac12
  \erf
  \left(
    \frac{t-t_M}{2\sqrt{t_M}}
  \right) .
\ee

Equation (\ref{R-Mott}) describes the crossover from $R=0$
to the uncorrelated value $R=1$ at the Mott scale.
This crossover is characterized by a sufficiently narrow
width $\sqrt{t_M}\ll t_M$.
The formula (\ref{R-Mott})
exactly coincides with the result for the strictly 1D geometry
obtained in Ref.~\onlinecite{GDP83}.


\subsubsection{Beyond the Mott scale, $t\gg t_M$}

Finally, we address the behavior at scales much larger than
the Mott length. Here $R\approx 1$, and one is interested
in the small correction $R(\omega,x)-1$.

In the region $t_M\ll t\ll t_M^2$, $A(\omega,x)$ reduces to $A^{(1)}(\omega,x)$
which is entirely determined by the saddle point $k_*^{(1)}$, and we get
\be
\label{R:tm<t<tmsq}
  R(\omega,x)
  =
  1
  -
  \frac{\pi^{9/2}t_M^3}{32 t^{7/2}}
  \exp
  \left(
    - \frac{(t-t_M)^2}{4t}
  \right)
  .
\ee

Equation (\ref{R:tm<t<tmsq}) breaks down at longest distances,
$t \gg t_M^2$. Here the contribution of the saddle points $k_*^{(n)}$
with $n>1$ become equally important, indicating that the replacement
(\ref{sum2int}) of the sum over discrete levels by an integral
is not adequate. Indeed, at $t \gg t_M^2$ the discrete energy
levels in the bosonic sector are completely resolved, see Eq.~(\ref{kn}).
The main correction to $R(\omega,x) = 1$ comes
from the ground state with $k \simeq \pi/\ln(1/4\Omega)$:
\be
  R(\omega,x)
  =
  1 + 8 \pi^3 \Re \frac{i k \tanh \pi k}{\cosh^2 \pi k}
    \frac{M_{0k}(\Omega) e^{-(k^2+1/4)t}}{\partial \ln S(k)/\partial k} .
\ee
The scattering matrix for such a small $k$ is
$S(k) \approx S_0(k) \approx e^{2i k \ln(1/4\Omega)}$.
Using the matrix element from Eq.\ (\ref{M0}) and
performing analytic continuation, we obtain
\begin{multline}
\label{R-oscillating}
  R(\omega,x)
  =
  1
  -
  \frac{2\pi^6}{t_M^3}
    \sin \left(
      \frac{8\pi^3 t_M t}{(t_M^2+\pi^2)^2}
    \right)
\\ {}
  \times
  \exp \left(
    - \frac{t}{4}
    + \frac{t_M}{2}
    - \frac{4\pi^2 (t_M^2-\pi^2) t}{(t_M^2+\pi^2)^2}
  \right)
  .
\end{multline}
In the region $t_M^2\ll t\ll t_M^4$, Eq.~(\ref{R-oscillating}) simplifies to
\be
\label{R-oscillating2}
  R(\omega,x)
  =
  1
  -
  \frac{2\pi^6}{t_M^3}
    \sin \left(
      \frac{8\pi^3 t}{t_M^3}
    \right) e^{-t/4+t_M/2-4\pi^2 t/t_M^2}
  .
\ee

Oscillations of the difference $R(\omega,x)-1$ with $x$
can be easily obtained in the opposite limit of large
frequencies, $\omega\gg\Delta_\xi$, where localization effects
are weak and the standard perturbation theory can be applied.
In this regime, $R(\omega,x)-1$ simultaneously decays and oscillates
with the characteristic scale given by the diffusive length
$\sqrt{D/\omega}$. At present, we do not know any explanation
of the fact that these oscillations persist well into the
localized region or of the physical origin
of the oscillation period scaling as $t_M^3$.

\begin{table}[t]
\caption{The poles of the integrand (\ref{An>0}) contributing
to the coefficient $a_1(t)$ in front of the $\omega^2$ term
in Eq.~(\ref{A-series}). The index $n$ labels the power of the
$S$-matrix, and the index $m$ labels the fermionic state
in the matrix-element block $M_{mk}$.}
\label{T:poles}
\begin{ruledtabular}
\begin{tabular}{cccc}
   pole   & $n$ & $m$ & exponent \\ \hline
   $i/2$  & 1   & 0   & 1 \\
          & 2   & 0   & 1 \\
          & 3   & 0   & 1 \\
          & 1   & 1   & $e^{-2t}$ \\ \hline
   $i$    & 1   & 0   & $e^{3t/4}$ \\ \hline
   $3i/2$ & 1   & 0   & $e^{2t}$
\end{tabular}
\end{ruledtabular}
\end{table}

\subsection{$\omega^2$ correction to $R(\omega,t)$}

Since to the leading order in small $\omega$
the result (\ref{R-decay}) coincides with its 1D analog,
it is instructive to study the subleading corrections
which would reveal the difference with the strictly 1D situation.
Such a difference has already been seen \cite{SO07} in the limit
of small spatial separations, $t\ll1$, and now we calculate
the correction to Eq.~(\ref{R-decay}) at arbitrary $t$.

At sufficiently small $t$, the function $A(\omega,t)$
can be formally expanded in powers of $\omega/\Delta_\xi$:
\be
  A(\omega,t)
  =
  \sum_{l=0}^\infty a_l(t) (\omega/\Delta_\xi)^{2l} ,
\label{A-series}
\ee
where the coefficients $a_l(t)$ are polynomials
of $\ln(\Delta_\xi/\omega)$ of order two.
The series (\ref{A-series}) is asymptotic:
at $t\sim t_M/(2l+1)$ the $l$'the term in the
sum acquires a non-analytic dependence on $\omega$
[the upturn (\ref{R-Mott}) at the Mott scale is an example
of such a behavior for $l=0$].

To the leading order we have
\be
  a_0(t) = -1 + A^{(0)}(0,t) ,
\label{a0(t)}
\ee
where $A^{(0)}(0,t)$ is given by Eq.~(\ref{A00}).

The subleading coefficient $a_1(t)$ can be obtained from the
general expression (\ref{A-M}) by adding the contributions
from $A^{(0)}(\omega,t)$ and $A^{(n>0)}(\omega,t)$.
The latter are calculated by deforming the integration
contour to the upper half plane and picking the residues
at the relevant poles specified in Table \ref{T:poles}.
The resulting expression has the form
\begin{widetext}
\begin{multline}
\label{a1(t)}
  a_1(t)
  =
  \left[
    \frac{(2 {\cal L}-3 t)^2}{24} + \frac{17 {\cal L}-24 t}{18}
  + \frac{829}{432} - \frac{\pi^2}{24}
  \right]
  e^{2 t}
- \frac{9\pi^4}{512} \, e^{3 t/4}
+ \frac{43}{72}
+ \left[
    \frac{(2 {\cal L}-t)^2}{24} + \frac{3 {\cal L}-t}{6}
    +\frac{7}{16} - \frac{\pi^2}{24}
  \right]
  e^{-2 t}
\\
  -
  \frac{\pi^2}{4}
  \int_0^\infty dk \, k \, \frac{\tanh\pi k}{\cosh^2\pi k}
      \left(
        \frac{(k^2+1/4)^2 (40k^4+56k^2+7)}{18(k^2+1)^2}
      + \frac{2(k^2+1/4)^3}{3(k^2+1)} \, t
      \right)
    e^{-(k^2+1/4)t}
  ,
\end{multline}
\end{widetext}
where
\be
\label{L-2gamma}
  {\cal L}
  =
  \ln(\Delta_\xi/\omega)
  -
  2\gamma
  =
  \frac{t_M}{2} - 2 \gamma ,
\ee
and $\gamma=0.577\dots$ is the Euler's constant.

Equation (\ref{a1(t)}) is valid for $t<t_M/3$ and provides the exact
distance dependence of the $\omega^2$ correction to the main
frequency-independent part $R(0,x)$ calculated in the previous section, see
Eq.\ (\ref{R-decay}). At $t\sim t_M/3$, the saddle point $k_*^{(1)}$ goes below
the pole $3i/2$, annihilating its contribution at $t>t_M/3$.
At longer distances, the leading frequency-dependent correction is due to the
saddle point in $A^{(1)}(\omega,t)$ and is no longer proportional to $\omega^2$.

\subsection{Summary}

In Table \ref{Tab:summary} we collect the results for various contributions to
$R(\omega,t)$ obtained in the previous Sections.
Besides that, the decay at the localization length scale
and the behavior at the Mott scale are given
by Eqs.~(\ref{R-decay}) and (\ref{R-Mott}), respectively.

\begin{table*}
\caption{Summary of the results for $R(\omega,x)$ at small frequencies and
arbitrary distances, $t=x/\xi$.}
\label{Tab:summary}
\begin{ruledtabular}
\begin{tabular}{ccc}
 Distance &
 $\omega$-independent term &
 $\omega$-dependent term ($t_M = 2\ln(\Delta_\xi/\omega)\gg1$)
\\\hline\\[-10pt]
 $1 \ll t < \dfrac{t_M}{3}$ &
 $\dfrac{\pi^{7/2}}{16 t^{3/2}} \exp\left(-\dfrac{t}{4}\right)$ &
 $\dfrac{\omega^2}{24\Delta_\xi^2}
   (t_M - 3t)^2 \exp\left( 2t \right)$
\\[10pt]
 $\dfrac{t_M}{3} < t < t_M$ &
 $\dfrac{\pi^{7/2}}{16 t^{3/2}} \exp\left(-\dfrac{t}{4}\right)$ &
 $\displaystyle \sim \exp \left( -\frac{(t - t_M)^2}{4t} \right)$
\\[10pt]
 $t_M \ll t \ll t_M^2$ &
 $1$ &
 $\displaystyle -\dfrac{\pi^{9/2}t_M^3}{32 t^{7/2}} \exp \left( -\frac{(t - t_M)^2}{4t} \right)$
\\[10pt]
 $t_M^2 \ll t$ &
 $1$ &
 $\displaystyle
  - \frac{2\pi^6}{t_M^3}
    \sin \left(
      \frac{8\pi^3 t_M t}{(t_M^2+\pi^2)^2}
    \right)
  \exp \left(
    - \frac{t}{4}
    + \frac{t_M}{2}
    - \frac{4\pi^2 (t_M^2-\pi^2) t}{(t_M^2+\pi^2)^2}
  \right)$
\end{tabular}
\end{ruledtabular}
\end{table*}

\section{Behavior of $R(\omega,t)$ at small $t$ and at arbitrary $\omega$}
\label{S:small-omega}

Here we present an alternative approach to calculating $A(\omega,t)$
which does not require the knowledge of the eigensystem
of the Hamiltonian $\tilde H$. This
approach applies to the case of small $t$ and arbitrary $\omega$ and
amounts to evaluating the expansion of $A(\omega,t)$ in powers of $t$.

We expand the evolution operator $e^{-2\tilde H t}$ in the general
expression
(\ref{A}) for $A(\omega,t)$ in series over $t$:
\be
\label{t-expansion}
   A(\omega,t)
   =
   \sum_{n=0}^\infty A_n(\omega) t^n,
\ee
where
\be
   A_n(\omega)
   =
   \frac12 \frac{(-1)^n}{n!} \Re \corr{\Psi_0|(2\tilde H)^n|\Psi_0}
  .
\ee

It is convenient to represent the matrix elements in the above expression in
terms of the variables $p$ and $q$ defined in Eq.\ (\ref{pq}). The
Hamiltonian
$\tilde H$, Eqs. (\ref{tilde-HB}) and (\ref{tilde-HF}), acquire the
following
form in this representation
\be
  \tilde H
   = \frac{1}{8} \left[
       \frac{1}{p} \partial_{p} p(\kappa^2-p^2) \partial_{p}
       +\frac{1}{q} \partial_{q} q(q^2-\kappa^2) \partial_{q}
       +p^2 - q^2
     \right],
\ee
where $\kappa^2 = 16 \Omega = -4i\omega/\Delta_\xi$. This operator is
defined
in the region $0 < q < \kappa$, $p > \kappa$.

The ground state wave function $\Psi_0$ [see Eq.\ (\ref{Psi0})] contains
modified Bessel functions of the arguments $p$ and $q$ with indices $0$
and $1$
only. The result of the action of $\tilde H$ on such a function can be
reduced
to a combination of the same Bessel functions multiplied by some
polynomials in
$p$ and $q$. We arrange the resulting expression in the form
\be
\label{H|0>}
   (2\tilde H)^n |\Psi_0\rangle
   =
   \sum_{i,k=0}^1 q^i p^k c_{ik}^{(n)}(p,q,\kappa) I_i(q) K_k(p)
   .
\ee
The matrices $c^{(n)}$ are even polynomials of $q$, $p$, and $\kappa$. Their
first entries are:
\begin{gather}
   c^{(0)} = \begin{pmatrix}
     0 & 1 \\
     1 & 0
   \end{pmatrix}
   ,
\nonumber
\\
   c^{(1)} = \begin{pmatrix}
     p^2+q^2-\kappa^2 & 0 \\
     0 & 1
   \end{pmatrix}
   ,
\\
   c^{(2)} = \begin{pmatrix}
     -2(p^2-q^2) & \frac32(p^2+q^2)-2\kappa^2 \\
     \frac32(p^2+q^2)-2\kappa^2 & 0
   \end{pmatrix}
   .
\nonumber
\end{gather}

To find $A_n(\omega)$, we have to multiply Eq.\ (\ref{H|0>}) by $\Psi_0$ and
integrate with $d\lambda_F = q\, dq/4\Omega$ and $d\lambda_B = p\,
dp/4\Omega$.
The integrals of Bessel functions arising in this calculation reduce to the
products of modified Bessel functions and polynomials, owing to the fact
that
the matrix $c^{(n)}$ contains only even powers of $p$ and $q$. Thus we can
calculate the functions $A_n(\omega)$ up to any $n$ recursively. The first
three terms of the expansion (\ref{t-expansion}) are
$A_0(\omega) = A(\omega,0)$ given by Eq.~(\ref{Aw0}),
$A_1(\omega) = B(\omega)$ given by Eq.~(\ref{B}),
and
\begin{align}
  A_2(\omega)
   &= \frac{4}{3} + \frac{2}{15} \Re \Big[
       (3\kappa^4 - 11 \kappa^2 + 16) (I_1^2 K_0^2 + I_0^2 K_1^2) \notag \\
       &+4\kappa(\kappa^2 - 4) (I_1 K_0 - I_0 K_1) (I_0 K_0 + I_1 K_1)
\notag \\
       &-3(\kappa^4 - 3 \kappa^2 + 4) I_1^2 K_1^2
       -\kappa^2 (3\kappa^2 - 8) I_0^2 K_0^2
     \Big],
\end{align}
where we omit the argument $\kappa$ of the Bessel functions for brevity.

In the limit $\omega \ll \Delta_\xi$, we can expand the
cumbersome exact expressions for $A_n(\omega)$ in powers of $\omega$.
The first two terms of this expansion are
[$\cal L$ is defined in Eq.~(\ref{L-2gamma})]:
\begin{align}
\label{A:A0}
  A_0(\omega)
   &= -\frac{1}{3} + \frac{\omega^2}{36 \Delta_\xi^2} \big(
       12 \mathcal{L}^2 + 52 \mathcal{L} + 43 - 3\pi^2
     \big), \\
\label{A:A1}
  A_1(\omega)
   &= -\frac{2}{3} + \frac{2 \omega^2}{9 \Delta_\xi^2}\big(
       \mathcal{L} + 1
     \big), \\
\label{A:A2}
  A_2(\omega)
   &= \frac{8}{15} + \frac{\omega^2}{18 \Delta_\xi^2} \big(
       12 \mathcal{L}^2 + 40 \mathcal{L} + 41 - 3\pi^2
     \big).
\end{align}
The same terms naturally appear in the expansion of $R(\omega,t)$, Eqs.\
(\ref{a0(t)}) and (\ref{a1(t)}), in powers of small $t$.

To make sure that the approach developed in this Section
is completely equivalent to the one adopted in the rest
of the paper, we have calculated the first 15 coefficients
$A_n(0)$ and verified that they coincide with the coefficients
obtained by expanding Eq.~(\ref{A00}) in a series in $t$.

\section{Discussion}
\label{S:Discussion}

In this work, we have developed a general approach to calculating
the correlation function of the LDOS in Q1D disordered wires
in the unitary symmetry class
using the quantum-mechanical reformulation
of the supersymmetric $\sigma$-model.
Our approach is convenient both for numerical
evaluation of the correlation function (see Fig.~\ref{F:Mott})
and for analytical treatment of various asymptotic regions.

To the leading order in small $\omega$, we demonstrate that
the decay of $R(\omega,x)$ at the localization length and its
jump to the uncorrelated value of 1
at the Mott scale are
described by exactly the same formulae in Q1D and strictly
1D geometries (with the natural
replacement $\xi\to\xi_\text{1D}$ and $L_M\to L_M^\text{1D}$).
However already the next-to-leading correction at small $\omega$
are different:\cite{SO07} $\omega^2\ln^2\omega$ in the Q1D geometry
and $\omega^2\ln\omega$ in the 1D geometry.

We believe that the difference between the subleading terms in the 1D
and Q1D problems should be attributed to the short-scale structure
of the wave functions. In this case, the $\omega$
dependence of the subleading terms should be sensitive to the symmetry
of the Q1D problem, with the orthogonal case being different from the
unitary case considered in the present work.

In the limit of very large spatial separations, $x\gg L_M$,
we identify a number of super-Mott scales.
At $x\sim \xi \ln^2(\Delta_\xi/\omega)$, the asymptotic
behavior of $R(\omega,x)$ changes its functional dependence,
leading to a faster decrease of correlations.
Even a larger super-Mott scale, $\xi \ln^3(\Delta_\xi/\omega)$,
gives the period of decaying oscillations in $R(\omega,x)-1$,
according to Eq.~(\ref{R-oscillating2}). The physical origin
of these super-Mott scales still remains to be clarified.
Note that the very-large-$x$ asymptotics of the LDOS correlator
in the 1D case has not been addressed in Ref.~\onlinecite{GDP83},
and it is thus impossible to make a comparison between the 1D
and Q1D results in this region.

The other interesting feature of the problem is the appearance
of the whole hierarchy of sub-Mott scales. Each time the saddle point
$k_*^{(n)}$ crosses a pole, a narrow step of the width
$\delta x\sim\xi\ln^{1/2}(\Delta_\xi/\omega)$ arises at some
rational value of $x/L_M$. The erf behavior
in Eq.~(\ref{R-Mott}), originating from the interplay
between the saddle $k_*^{(1)}$ and the pole $i/2$ is the strongest of
those steps appearing already in the leading order.
The amplitudes of other, sub-Mott, steps are proportional
to some powers of $\omega/\Delta_\xi$ and therefore they can
be seen only in subleading contributions.

We believe that our technique developed for calculating
the LDOS correlation functions can be used to find
other low-frequency properties of Q1D systems.
The long-standing problem in the field concerns the derivation
of the low-frequency dissipative conductivity in Q1D wires.
The main technical difficulty compared to the case of the LDOS
correlations is that calculation of conductivity involves
a much more complicated evolution operator \cite{Efetov-book,DM}
compared to (\ref{A}).

Finally, it remains a challenging task to generalize our approach
from the unitary to other symmetry classes. This is a sophisticated
problem since the corresponding quantum-mechanical formulation
involves more than two independent variables.\cite{Efetov-book}


\acknowledgments

We thank D. N. Aristov, Y. V. Fyodorov, I. V. Gornyi, V. E. Kravtsov, and A. D.
Mirlin
for stimulating discussions.
P.~O. and M.~S. acknowledge the hospitality
of the Institute for Theoretical Physics at EPFL,
where the main part of this project was done.
The work by P.~O. and M.~S. was partially supported
by the RFBR grant No.\ 07-02-00976.

\appendix

\section{Normalization of the eigenfunction $\phi_k(\lambda)$}
\label{A:normalization}

Here we compute the norm of the function $\phi_k(\lambda)$ given by Eq.\
(\ref{phik}):
\be
 \langle \phi_k | \phi_k \rangle =
  \int_1^{\lambda^*} [L_k(\lambda)]^2 d\lambda
   + C_k^2 \int_{\lambda^*}^\infty [B_k(\lambda)]^2 d\lambda
  ,
\label{normalization-split}
\ee
where $\lambda^*$ is an arbitrarily chosen value in the
intermediate-asymptotics region $1\ll \lambda^* \ll 1/\Omega$.

We can expand each of the two integrals in Eq.\
(\ref{normalization-split}) as a
power series in $(\lambda^*+1)$, and
their sum should be independent of $\lambda^*$. So we can keep only
zeroth-order terms in $(\lambda^*+1)$ from the start, to simplify the
calculation. We denote the operation of singling out zeroth-order terms
(constants and logarithms) by $\ZO$.

To compute the normalization in a systematic way, we expand
$L_k(\lambda)$ and $B_k(\lambda)$ via (\ref{L-series}) and (\ref{B-series})
and combine the terms as
\be
  \langle \phi_k | \phi_k \rangle =
  \sum_{n,m=-\infty}^\infty c_n c_m I_{nm}
  ,
\label{normalization-sum}
\ee
where in
\be
  I_{nm}=I_{nm}^B + I_{nm}^L
\ee
we only keep zeroth-order [constants
or logarithms] terms in $(\lambda^*+1)$ in the integrals
\begin{align}
 I_{nm}^B & = C_k^2 \ZO \int_{\lambda^*}^\infty
 B^{(0)}_{k+in} (\lambda) B^{(0)}_{k+im} (\lambda) \, d\lambda
 ,
\label{Inm-Bessel}
\\
 I_{nm}^L & = \ZO \int_1^{\lambda^*}
 L^{(0)}_{k+in} (\lambda) L^{(0)}_{k+im} (\lambda) \, d\lambda
 .
\label{Inm-Legendre}
\end{align}
There are two types of such terms: diagonal ($n=m$) and
off-diagonal ($n\ne m$). They have different structure and are
computed by different methods.

\subsection{Diagonal terms}

Diagonal terms give rise to contributions scaling as $\log\Omega$.
They are computed with the use of the normalization lemma in
Appendix \ref{norm-lemma}.

For the case of Bessel functions, the integral (\ref{Inm-Bessel})
at $n=m=0$ is written as
\begin{multline}
\ZO \int_{\lambda^*}^\infty [B^{(0)}_k]^2\, d\lambda
\\ =
\int_{-1+\epsilon}^\infty [B^{(0)}_k]^2\, d\lambda -
\ZO \int_{-1+\epsilon}^{\lambda^*} [B^{(0)}_k]^2\, d\lambda
  .
\label{diagonal-Bessel}
\end{multline}
We imply the limit $\epsilon\to0$ so that the regular part (with oscillating
terms ommited) of the first
term may be calculated using the normalization lemma in Appendix
\ref{norm-lemma}:
\begin{multline}
\label{reg-bessel1}
\Reg \int_{-1+\epsilon}^\infty [B^{(0)}_k]^2\, d\lambda
\\ =
\frac{\Gamma(2ik)\Gamma(-2ik)}{16\Omega}
\left[ -\ln (2\Omega\epsilon) + \frac{1}{2i}
\frac{\partial}{\partial k} \ln \frac{\Gamma(2ik)}{\Gamma(-2ik)}
\right]
  .
\end{multline}
The second integral in (\ref{diagonal-Bessel}) is calculated
using the expansion of the Bessel functions (\ref{B-hypergeometric}) at
$\Omega(\lambda+1) \ll 1$.
We again omit all oscillating terms and obtain
\begin{multline}
\label{reg-bessel2}
\ZO \Reg
\int_{-1+\epsilon}^{\lambda^*} [B^{(0)}_k]^2\, d\lambda
\\ =
\frac{\Gamma(2ik)\Gamma(-2ik)}{16\Omega}
\ln\frac{\lambda^*+1}{\epsilon}
  .
\end{multline}

Adding the two contributions (\ref{reg-bessel1}) and (\ref{reg-bessel2}) cancels
the dependence on $\epsilon$ as it should be and gives
\begin{multline}
I^B_{00}= C_k^2
\frac{\Gamma(2ik)\Gamma(-2ik)}{16\Omega}
\\ \times
\left[ -\ln[2\Omega(\lambda^*+1)] +
\frac{1}{2i} \frac{\partial}{\partial k}
\ln\frac{\Gamma(2ik)}{\Gamma(-2ik)}\right]
  .
\end{multline}

We repeat the same procedure for the Legendre part:
\begin{multline}
\ZO \int_{1}^{\lambda^*} [L^{(0)}_k]^2\, d\lambda
\\ =
\int_{1}^{\Lambda} [L^{(0)}_k]^2\, d\lambda -
\ZO \int_{\lambda^*}^{\Lambda} [L^{(0)}_k]^2\, d\lambda
  ,
\label{diagonal-Legendre}
\end{multline}
with $\Lambda \to \infty$.
Applying the normalization lemma from Appendix
\ref{norm-lemma}, we find the regular (without oscillations) part of the first
term:
\begin{multline}
\Reg \int_{1}^{\Lambda} [L^{(0)}_k]^2\, d\lambda =
\frac{\Gamma(2ik)\Gamma(-2ik)}{\Gamma^2(ik+1/2)\Gamma^2(-ik+1/2)}
\\ \times
4 \left[ \ln\frac{\Lambda+1}{2} + \frac{1}{2i} \frac{\partial}{\partial k}
\ln\frac{\Gamma(2ik)\Gamma^2(-ik+1/2)}{\Gamma(-2ik)\Gamma^2(ik+1/2)} \right]
  .
\end{multline}
In calculating the regularized second term of Eq.\ (\ref{diagonal-Legendre}) we
use the expansion (\ref{L-hypergeometric}) and obtain
\begin{multline}
\ZO \Reg
\int_{\lambda^*}^{\Lambda} [L^{(0)}_k]^2\, d\lambda
\\ =
\frac{\Gamma(2ik)\Gamma(-2ik)}{\Gamma^2(ik+1/2)\Gamma^2(-ik+1/2)}
\cdot 4 \ln\frac{\Lambda+1}{\lambda^*+1}
  .
\end{multline}
Adding the two terms, we find
\begin{multline}
I^L_{00}=
\frac{\Gamma(2ik)\Gamma(-2ik)}{\Gamma^2(ik+1/2)\Gamma^2(-ik+1/2)}
\\ \times
4 \left[ \ln\frac{\lambda^*+1}{2} + \frac{1}{2i} \frac{\partial}{\partial k}
\ln\frac{\Gamma(2ik)\Gamma^2(-ik+1/2)}{\Gamma(-2ik)\Gamma^2(ik+1/2)} \right]
  .
\end{multline}

Other diagonal elements $I^B_{nn}$ and $I^L_{nn}$ may be obtained by
analytically continuing  $I^B_{00}$ and $I^L_{00}$ to $k\to k+in$.
Note that such an analytic continuation commutes with the $\ZO$ operation.

Finally, we add the Bessel and Legendre contributions together and,
using the identity
\be
\frac{\Gamma(2ik)\Gamma(-2ik)}{\Gamma^2(ik+1/2) \Gamma^2(-ik+1/2)}
= \frac{\coth (\pi k)}{4\pi k}\, ,
\label{identity-coth}
\ee
obtain
\be
I_{00}(k)=
\frac{\coth(\pi k)}{2\pi i k} \frac{\partial}{\partial k}
\ln S_0(k)
  ,
\ee
where $S_0(k)$ is the scattering matrix (\ref{S0(k)})
without corrections in $\Omega$.

The other diagonal terms are given by the analytic continuation
\be
I_{nn}(k)=
I_{00}(k+in)
= \frac{\coth(\pi k)}{2\pi i (k+in)} \frac{\partial}{\partial k}
\ln S_0(k+in)
  .
\label{Inn}
\ee

\subsection{Off-diagonal terms}

Off-diagonal terms form a series in $\Omega$ with positive powers only.
They involve overlaps of two functions which solve the same Hamiltonian
at different (complex) energies. Such integrals may be conveniently
computed by the Wronskian method, which results in
\begin{multline}
I_{nm}^B = C_k^2 \ZO
\frac{(\lambda^*+1)^2}{(k+in)^2-(k+im)^2}
\\ \times
\left[
\partial_{\lambda^*} B^{(0)}_{k+in} (\lambda^*) B^{(0)}_{k+im} (\lambda^*)
-
B^{(0)}_{k+in} (\lambda^*) \partial_{\lambda^*} B^{(0)}_{k+im} (\lambda^*)
\right]
\end{multline}
and
\begin{multline}
I_{nm}^L = - \ZO
\frac{(\lambda^*+1)(\lambda^*-1)}{(k+in)^2-(k+im)^2}
\\ \times
\left[
\partial_{\lambda^*} L^{(0)}_{k+in} (\lambda^*) L^{(0)}_{k+im} (\lambda^*)
-
L^{(0)}_{k+in} (\lambda^*) \partial_{\lambda^*} L^{(0)}_{k+im} (\lambda^*)
\right]
  .
\end{multline}

These coefficients have the obvious symmetries
\be
 I_{mn}=I_{nm}=I_{-n,-m}(k \to -k) .
\ee

By using the expansions (\ref{L-hypergeometric}), (\ref{B-hypergeometric})
of the Legendre and MacDonald functions, after some tedious
combinatorics, one obtains the following form of the off-diagonal
coefficients (we have used the relation (\ref{identity-coth})
to simplify products of Gamma functions):
\be
  I^B_{nm}
  = (-1)^{n+m} I^L_{nm}
  = \frac{\coth(\pi k)}{\pi |n-m| \left(k+i\frac{n+m}{2} \right)}
  ,
\label{Inm-1}
\ee
which leads to
\be
  I_{nm}
  =
  \begin{cases}
    0, & m+n \equiv 1 \pmod 2 , \\
    \dfrac{2\coth(\pi k)}{\pi |n-m| \left(k+i\frac{n+m}{2} \right)},
      & m+n \equiv 0 \pmod 2 .
  \end{cases}
\ee

Thus the Legendre and Bessel contributions cancel each other
at odd powers of $\Omega$ and double each other at even powers of $\Omega$.

\subsection{Normalization}

The diagonal terms (\ref{Inn})
in the sum (\ref{normalization-sum}) contain logarithmic
derivatives of $S_0(k + i n)$ taken at different $n$.
We may perform two transformation on those terms. First,
with the help of gamma-function properties,
we reduce the argument of $S_0(k + i n)$ to $k$ acquiring
new terms which form a series in $\Omega$.
Second, we replace $\ln S_0(k)$ by $\ln S(k)$ according
to Eq.~(\ref{S(k)}), again at the cost of adding a
series in $\Omega$.
Surprisingly, these two contributions exactly cancel the
off-diagonal part of the sum (\ref{normalization-sum})!
We have observed this cancellation up to the order $O(\Omega^6)$
and conjecture that it is exact at all orders.


The resulting expression (after canceling the off-diagonal terms)
is proportional to $\partial \ln S(k) / \partial k$
with the full $S$-matrix (\ref{S(k)}):
\begin{equation}
\langle \phi_k | \phi_k \rangle
= -i\frac{\coth \pi k}{2\pi} \frac{\partial \ln S(k)}{\partial k}
\sum_{n=-\infty}^\infty \frac{[c_n(\Omega,k)]^2}{k + i n}.
\end{equation}

\section{Normalization lemma}
\label{norm-lemma}

In one dimension, consider a particle moving freely in one
asymptotic direction and fully reflecting of a potential in
the other direction.
Then the normalization of the wave function in the region
where the particle experiences reflection may be related to
the scattering phase shift.

Specifically, consider the one-dimensional Hamiltonian
\begin{equation}
H=\frac{p^2}{2m} + U(x)
\end{equation}
with the condition on the potential $U(x)=0$ for $x<0$ and
$U(x)\to\infty$ at $x\to\infty$. Then to any positive momentum $k$
there corresponds a state of the continuous spectrum $\Psi_k(x)$
with the asymptotic behavior at $x<0$:
\begin{equation}
\Psi_k(x<0) = e^{ik(x-x_0)} - e^{-ik(x-x_0)+i \varphi_{x_0}(k)} ,
\end{equation}
where $x_0$ is an arbitrarily chosen reference point, and $\varphi_{x_0}(k)$
is the scattering phase relative to this point. Obviously, $\varphi_{x_0}(k)$
is a linear function of $x_0$ with the slope $2k$. Consider now the
normalization of the part of the wave function from $x_0$ to $+\infty$:
\begin{equation}
I_{x_0,k} = \int_{x_0}^\infty |\Psi_k(x)|^2 dx .
\end{equation}
Obviously, for $x_0<0$ this normalization is a linear function of $x_0$ plus
a sinusoidal function. We define the {\it regularized} normalization integral
\begin{equation}
\Reg I_{x_0,k} =\Reg \int_{x_0}^\infty |\Psi_k(x)|^2 dx
\end{equation}
as $I_{x_0,k}$ with the sinusoidal part subtracted.

Then our normalization lemma states that
\begin{equation}
\Reg I_{x_0,k} = \frac{\partial \varphi_{x_0}(k)}{\partial k} .
\label{lemma-result}
\end{equation}

To prove the lemma, consider the same problem with an infinite
wall placed at a position $x_0<0$. Then, for any two points
$x,x' > x_0$ we have two decompositions of the delta function:
the one by the full basis of the continuous spectrum,
\begin{equation}
\int \frac{dk}{2\pi} \Psi_k^*(x) \Psi_k(x') = \delta(x-x') ,
\label{lemma-delta-1}
\end{equation}
and the other one by the basis of the discrete spectrum of the
states constrained by the infinite wall at $x_0$:
\begin{equation}
\sum_k  ( I_{x_0,k} )^{-1} \Psi_k^*(x) \Psi_k(x') =  \delta(x-x') , \qquad
x,x'>x_0 .
\label{lemma-delta-2}
\end{equation}
First, note that in the case of an infinite wall, the normalization of the
states of the discrete spectrum coincides with the regularized one:
$I_{x_0,k} = \Reg I_{x_0,k}$. Second, we may replace
\begin{equation}
\sum_k \to \int \frac{dk}{2\pi} \sum_{n=-\infty}^{+\infty}
e^{in\varphi_{x_0}(k)} \frac{\partial \varphi_{x_0}(k)}{\partial k}
\end{equation}
and after subtracting (\ref{lemma-delta-1}) from (\ref{lemma-delta-2})
we obtain
\begin{multline}
\int \frac{dk}{2\pi} \left[\frac{\partial \varphi_{x_0}(k)}{\partial k}
( \Reg I_{x_0,k} )^{-1} -1 \right] \Psi_k^*(x) \Psi_k(x') \\
 + \sum_{n\ne 0}
\int \frac{dk}{2\pi} e^{in\varphi_{x_0}(k)}
\frac{\partial \varphi_{x_0}(k)}{\partial k} (\Reg I_{x_0,k} )^{-1}
\Psi_k^*(x) \Psi_k(x') \\
=0 \quad \text{for any $x,x'>x_0$}.
\end{multline}
Now we take the limit $x_0\to -\infty$. In this limit, the first
term goes to zero as $1/|x_0|$, since both
$\partial \varphi_{x_0}(k)/\partial k$ and $\Reg I_{x_0,k}$ are linear
functions of $x_0$. The second term goes to zero faster than $1/|x_0|$,
because of the rapidly oscillating exponent. Therefore, the first term must
identically vanish, which implies the result (\ref{lemma-result}).

\section{Evaluation of $\corr{K_0|\phi_k}$ and $\corr{p K_1|\phi_k}$}
\label{A:overlap}

Let us start with the matrix element $\corr{K_0|\phi_k}$. Using the
perturbative expansions (\ref{L-series}) and (\ref{B-series}) of the
wave function $\phi_k$, we represent the matrix element in the form
\begin{multline}
 \corr{K_0|\phi_k}
  = \sum_{n=0}^\infty c_n(\Omega, k)\bigg[
      \int_1^{\lambda^*} d\lambda\; L^{(0)}_{k+in}(\lambda) K_0(p) \\
      + C_k \int_{\lambda^*}^\infty d\lambda\; B^{(0)}_{k+in}(\lambda)
K_0(p)
    \bigg]
\label{K0phi-integral}
\end{multline}
with $1 \ll \lambda^* \ll 1/\Omega$. We then use the hypergeometric
series (\ref{L-hypergeometric}), (\ref{B-hypergeometric}) for Legendre
and Bessel functions and integrate every term of these expansions with
$K_0(p)$ with the help of the following indefinite integral
\begin{multline}
 \int dp\; p^a K_0(p)
  = \frac{p^{1+a} K_0(p)}{1+a}\, {}_1F_2 \left(
      1; \frac{1+a}{2},\frac{3+a}{2}; \frac{p^2}{4}
    \right) \\
    + \frac{p^{2+a} K_1(p)}{(1+a)^2}\, {}_1F_2 \left(
      1; \frac{3+a}{2},\frac{3+a}{2}; \frac{p^2}{4}
    \right).
\label{indefinite-K-integral}
\end{multline}
The parameter $k$ satisfies the quantization condition $S(k) = 1$, which
ensures that the two expansions of the wave function in Eq.\
(\ref{K0phi-integral}) match at the point $\lambda^*$. This allows us to
omit the dependence of the two integrals in Eq.\ (\ref{K0phi-integral})
on $\lambda^*$ by applying $\ZO$ operation defined in Appendix
\ref{A:normalization}.

Let us start with the Legendre part of Eq.\ (\ref{K0phi-integral}). We
integrate every term of the series (\ref{L-hypergeometric}) for
$L^{(0)}_k$ with $K_0(p)$ using Eq.\ (\ref{indefinite-K-integral}) and
expand the emerging hypergeometric functions. The contribution of the
upper limit $\lambda^*$ does not contain any constant or logrithmic
terms in $\lambda^*+1$ and therefore is totally annihilated by the $\ZO$
operation. At the lower limit $\lambda = 1$, we have
\begin{multline}
 \ZO \int_1^{\lambda^*} d\lambda\; L^{(0)}_k(\lambda) K_0(p) \\
  = \sum_{m,l=0}^\infty \frac{\Gamma(2ik - m)\; (-1)^{m+1} (4\Omega)^l}
    {\Gamma^2 \big( \frac{1}{2} + ik - m + l \big)\; m!} \Bigg[
      \frac{2 K_0(4\sqrt{\Omega})}{\frac{1}{2} - ik + m - l} \\
      +\frac{4\sqrt{\Omega} K_1(4\sqrt{\Omega})}
      {\big( \frac{1}{2} - ik + m - l \big)^2}
    \Bigg] + \{ k \mapsto -k \}.
\end{multline}

We first sum over the index $m$ and obtain
\begin{multline}
 \ZO \int_1^{\lambda^*} d\lambda\; L^{(0)}_k(\lambda) K_0(p) \\
  = \sum_{l=0}^\infty \frac{i \pi (2l)! (4\Omega)^l}
    {\sinh(2\pi k) \Gamma^2 \big( \frac{1}{2} + ik + l \big)
    \Gamma^2 \big( \frac{1}{2} - ik + l \big)} \\
  \times \left\{
      \frac{2 K_0(4\sqrt{\Omega})}{\big( \frac{1}{2} + n \big)^2 + k^2}
      +\frac{(2l+1)4\sqrt{\Omega}K_1(4\sqrt{\Omega})}
      {\big[ \big( \frac{1}{2} + n \big)^2 + k^2 \big]^2}
    \right\} \\
    + \{ k \mapsto -k \}.
\end{multline}
Every term in the above sum over $l$ is an odd function of $k$ and
cancels against its $(k \mapsto -k)$ counterpart. This immediately
yields zero:
\begin{equation}
 \ZO \int_1^{\lambda^*} d\lambda\; L^{(0)}_k(\lambda) K_0(p)
  = 0.
\label{RegL=0}
\end{equation}
This identity also holds after the shift of the parameter $k \mapsto k +
in$ because the integral remains convergent in the limit $\lambda \to 1$
and the operation $\ZO$ commutes with the imaginary momentum shift.

Now we turn to the second integral in Eq.\ (\ref{K0phi-integral}) and
rewrite it in the form
\begin{multline}
\label{B=B-B}
 \ZO \int_{\lambda^*}^\infty d\lambda\; B^{(0)}_k(\lambda) K_0(p)
  = \int_{-1}^\infty d\lambda\; B^{(0)}_k(\lambda) K_0(p) \\
    - \ZO \int_{-1}^{\lambda^*} d\lambda\; B^{(0)}_k(\lambda) K_0(p).
\end{multline}
The first integral (running from $-1$ to $\infty$) is known to be
\begin{equation}
 \int_{-1}^\infty d\lambda\; B^{(0)}_k(\lambda) K_0(p)
  = \frac{\pi^2}{16 \Omega \cosh^2 (\pi k)}.
\end{equation}
In the second integral we will omit the dependence on $\lambda^*$
according to the $\ZO$ prescription. Substituting the expansion of the
Bessel function (\ref{B-hypergeometric}) and using Eq.\
(\ref{indefinite-K-integral}), we see that every term of the
hypergeometric series yields zero at the lower limit $p=0$. Therefore
the Bessel part of the matrix element is
\begin{equation}
 \ZO \int_{\lambda^*}^\infty d\lambda\; B^{(0)}_k(\lambda) K_0(p)
  = \frac{\pi^2}{16 \Omega \cosh^2 (\pi k)}.
\end{equation}
This result holds for the shifted index $k \mapsto k + in$ as well,
despite the fact that both integrals in the right hand side of Eq.\
(\ref{B=B-B}) converge only when $k$ is real.

Collecting all the terms we obtain the matrix element
\begin{equation}
\corr{K_0|\phi_k}
  = \frac{\pi^2 C_k}{16 \Omega \cosh^2 (\pi k)}
    \sum_{n=0}^\infty c_n(\Omega, k).
\end{equation}

Calculation of $\corr{p K_1|\phi_k}$ is very similar. We separate the
Legendre and Bessel parts of the matrix element as in Eq.\
(\ref{K0phi-integral}). With the help of the result (\ref{RegL=0}) we
again find that the Legendre integral yields zero
\begin{multline}
 \ZO \int_1^{\lambda^*} d\lambda\; L^{(0)}_k(\lambda) p K_1(p) \\
  = -2\Omega \frac{\partial}{\partial\Omega} \ZO
    \int_1^{\lambda^*} d\lambda\; L^{(0)}_k(\lambda) K_0(p)
  = 0.
\end{multline}
Next, we extend the Bessel integral to the whole axis, as we did in Eq.\
(\ref{B=B-B}), and find
\begin{multline}
 \ZO \int_{\lambda^*}^\infty d\lambda\; B^{(0)}_k(\lambda) p K_1(p)
  = \int_{-1}^\infty d\lambda\; B^{(0)}_k(\lambda) p K_1(p) \\
  = \frac{\pi^2 (1 + 4 k^2)}{32 \Omega \cosh^2 (\pi k)}.
\end{multline}
After the analytic continuation $k \mapsto k + in$ and collecting all the
terms, we get
\begin{equation}
 \corr{p K_1|\phi_k}
  = \frac{\pi^2 C_k}{32 \Omega \cosh^2 (\pi k)}
    \sum_{n=0}^\infty c_n(\Omega, k) \big[ 1 + 4(k + in)^2 \big].
\end{equation}

Using the constant $C_k$ from Eq.\ (\ref{C_k}) we finally obtain the
matrix elements in the form Eqs.\ (\ref{K0phik}) and (\ref{K1phik}).

\end{document}